\documentstyle[12pt,psfig]{article}
\def\baselinestretch{1.5}
\begin{document}
\pagenumbering{arabic}

\begin{center}
{\Large \bf Spectra of anti-nucleons in the vacuum of finite nuclei}\\
 \vspace{0.5cm}
Guangjun~Mao \\
\vspace{1.0cm}
{\em $^{1)}$Institute of High Energy Physics, Chinese Academy of Science \\
      P.O. Box 918(4), Beijing 100039, P.R. China\footnote{mailing address\\
      \indent $\;\;$ e-mail: maogj@mail.ihep.ac.cn}\\
$^{2)}$Institute of Theoretical Physics, Chinese Academy of Science\\
      P.O. Box 2735, Beijing 100080, P.R. China \\
$^{3)}$CCAST (World Lab.), P.O. Box 8730, Beijing 100080, P.R.
       China }
\end{center}
\date{\today}
\vspace{0.5cm}
\begin{abstract}
\begin{sloppypar}
The quantum vacuum in a many-body system of finite nuclei has been
investigated within the relativistic Hartree approach which
describes the bound states of nucleons and anti-nucleons
consistently. The contributions of the Dirac sea to the source
terms of the meson-field equations are taken into account up to
the one-nucleon loop and one-meson loop. The tensor couplings for
the $\omega$- and $\rho$-meson are included in the model. After
adjusting the parameters of the model to the properties of
spherical nuclei, a large effective nucleon mass $m^{*}/M_{N}
\approx 0.78$ is obtained. The overall nucleon spectra of
shell-model states are in agreement with the data. The computed
anti-nucleon spectra in the vacuum differ about 20 -- 30 MeV with
and without the tensor-coupling effects.

\end{sloppypar}
\end{abstract}
\newpage
\begin{center}
{\bf I. INTRODUCTION}
\end{center}
\begin{sloppypar}
One of the main characters distinguishing  relativistic approaches
from  nonrelativistic approaches is that the former one has a
vacuum. It is quite interesting to study the structure of quantum
vacuum in a many-body system, e.g., in a finite nucleus where the
Fermi sea is filled with the valence nucleons while the Dirac sea
is full of the virtual nucleon--anti-nucleon pairs. A schematic
picture taken from Refs. \cite{Aue86,Rei86} is given in Fig.~1.
The shell-model states have been theoretically and experimentally
well established \cite{Boh69} while no information for the bound
states of anti-nucleons in the Dirac sea are available. This is
the aim of our work. The observation of anti-nucleon bound states
is a verification for the application of the relativistic quantum
field theory  to a many-body system \cite{Ser86}. It constitutes a
basis for the widely used relativistic mean-field (RMF) theory
under the no-sea approximation
\cite{Rei86,Ser86,Gam90,Ren96,Men98,Sug94} and the relativistic
Hartree approach (RHA)\cite{Hor84,Per86,Was88,Fox89,Fur89}. Since
the bound states of nucleons are subject to the cancellation of
two potentials $S+V$ ($V$ is positive, $S$ is negative) while the
bound states of anti-nucleons, due to the G-parity, are sensitive
to the sum of them $S-V$, consistent studies of both the nucleon
and the anti-nucleon bound states can determine the individual $S$
and $V$. In addition, the exact knowledge of potential depth for
anti-nucleons in the medium is a prerequisite for the study of
anti-matter and anti-nuclei in relativistic heavy-ion collisions
\cite{Bea00,Arm00}.

\begin{center}
\fbox{Fig.~1}
\end{center}

We have developed a relativistic Hartree approach which describes
the bound states of nucleons and anti-nucleons in a unified
framework. For the details of the model we refer to Refs.
\cite{Mao99,Mao03}. A brief description will be given in Sec. II.
Numerical results and discussions are presented in Sec. III.

\end{sloppypar}
\begin{center}
{\bf II. RELATIVISTIC HARTREE APPROACH}
\end{center}
\begin{sloppypar}
The Lagrangian density of nucleons interacting through the
exchange of mesons can be expressed as \cite{Ser86}
\begin{eqnarray}
{\cal L}&=&\bar{\psi}[i\gamma_{\mu}\partial^{\mu}-M_{N}]\psi
   + \frac{1}{2}
\partial_{\mu}\sigma\partial^{\mu}\sigma-U(\sigma)
 -\frac{1}{4}\omega_{\mu\nu}\omega^{\mu\nu} \nonumber \\
&& + \frac{1}{2}m_{\omega}^{2}\omega_{\mu}\omega^{\mu}
 - \frac{1}{4} {\bf R}_{\mu\nu} \cdot {\bf R}^{\mu\nu}
 +\frac{1}{2}m_{\rho}^{2}{\bf R}_{\mu} \cdot {\bf R}^{\mu}
- \frac{1}{4}  A_{\mu\nu} A^{\mu\nu} \nonumber \\
      &&+ {\rm g}_{\sigma}\bar{\psi}\psi\sigma
      - {\rm g}_{\omega}\bar{\psi}\gamma_{\mu}\psi\omega^{\mu}
      -\frac{f_{\omega}}{4M_{N}}\bar{\psi}\sigma^{\mu\nu}\psi\omega_{\mu\nu}
  - \frac{1}{2}{\rm g}_{\rho}\bar{\psi}\gamma_{\mu}\mbox{\boldmath $\tau$}
    \cdot \psi {\bf R}^{\mu} \nonumber \\
 && -\frac{f_{\rho}}{8M_{N}}\bar{\psi}\sigma^{\mu\nu}\mbox{\boldmath
$\tau$}\cdot \psi {\bf R}_{\mu\nu} - \frac{1}{2}
e\bar{\psi}(1+\tau_{0})\gamma_{\mu} \psi A^{\mu},
    \end{eqnarray}
where U($\sigma$) is the self-interaction part of the scalar field
\cite{Bog77}
\begin{equation}
  U(\sigma)=
   \frac{1}{2}m_{\sigma}^{2}\sigma^{2}+\frac{1}{3!}b
\sigma^{3}+\frac{1}{4!}c\sigma^{4}.
\end{equation}
In the above $f_{\omega}$ and $f_{\rho}$ are the tensor-coupling
strengths of vector mesons; other symbols have their usual
meaning.

In finite nuclei the Dirac equation is written as
 \begin{eqnarray}
 i \frac{\partial}{\partial t}\psi({\bf x},t) &=&
        \left[ -i \mbox{\boldmath $\alpha$} \cdot \mbox{\boldmath $\nabla$}
        + \beta\left(M_{N} - {\rm g}_{\sigma}\sigma({\bf x})\right)
        +{\rm g}_{\omega}\omega_{0}({\bf x})
        -\frac{f_{\omega}}{2M_{N}}i\mbox{\boldmath $\gamma$}\cdot
        \left(\mbox{\boldmath $\nabla$} \omega_{0}({\bf x})\right)
        \right. \nonumber \\
    &+& \left. \frac{1}{2}{\rm g}_{\rho}\tau_{0}R_{0,0}({\bf x})
        -\frac{f_{\rho}}{4M_{N}}i\tau_{0}\mbox{\boldmath $\gamma$}\cdot
        \left(\mbox{\boldmath $\nabla$}R_{0,0}({\bf x})\right)
        +\frac{1}{2}e(1+\tau_{0})A_{0}({\bf x})\right] \psi({\bf x},t).
        \label{field}
 \end{eqnarray}
The field operator can be expanded according to nucleons and
anti-nucleons and reads as
 \begin{equation}
\psi({\bf x},t)=\sum_{\alpha} \left[ b_{\alpha}\psi_{\alpha}({\bf
 x}) e^{-i E_{\alpha}t} + d^{+}_{\alpha}\psi^{a}_{\alpha}({\bf x})
 e^{i \bar{E} _{\alpha} t} \right]. \label{opera}
  \end{equation}
Here the label $\alpha$ denotes the full set of single-particle
quantum numbers.  The wave functions of nucleons and anti-nucleons
can be specified as \cite{Bjo64,Mao99}
  \begin{equation}
 \psi_{\alpha}({\bf x})= \left( \begin{array}{l}
 i \frac{G_{\alpha}(r)}{r} \Omega_{jlm}(\frac{{\bf r}}{r})  \\
  \frac{F_{\alpha}(r)}{r}\frac{\mbox{\boldmath $\sigma$}\cdot {\bf r}}{r}
 \Omega_{jlm}(\frac{{\bf r}}{r}) \end{array} \right), \label{wfn}
  \end{equation}
  \begin{equation}
 \psi_{\alpha}^{a}({\bf x})= \left( \begin{array}{l}
- \frac{\bar{F}_{\alpha}(r)}{r}\frac{\mbox{\boldmath
$\sigma$}\cdot {\bf r}}{r}
 \Omega_{jlm}(\frac{{\bf r}}{r})  \\
  i \frac{\bar{G}_{\alpha}(r)}{r} \Omega_{jlm}(\frac{{\bf r}}{r})
 \end{array} \right). \label{wfan}
  \end{equation}
Here $\Omega_{jlm}$ are the spherical spinors.

Inserting Eq. (\ref{opera}) into Eq. (\ref{field}) and making some
straightforward algebra we arrive at the
Schr\"{o}dinger-equivalent  equations for the upper component of
the nucleon's  wave function
 \begin{equation}
E_{\alpha}G_{\alpha}(r) = \left[- \frac{d}{dr} + W(r)
 \right] M_{eff}^{-1} \left[ \frac{d}{dr} + W(r) \right]
 G_{\alpha}(r) + U_{eff}G_{\alpha}(r),  \label{schn}
 \end{equation}
and the lower component of the anti-nucleon's wave function
 \begin{equation}
\bar{E}_{\alpha}\bar{G}_{\alpha}(r) = \left[-
\frac{d}{dr}+\bar{W}(r) \right] \bar{M}_{eff}^{-1} \left[
\frac{d}{dr} + \bar{W}(r) \right]
 \bar{G}_{\alpha}(r) + \bar{U}_{eff}\bar{G}_{\alpha}(r).
 \label{schan}
 \end{equation}
Other components can be obtained through the following relations
 \begin{eqnarray}
&& F_{\alpha}(r) = M_{eff}^{-1} \left[ \frac{d}{dr} + W(r)
   \right] G_{\alpha}(r), \\
&& \bar{F}_{\alpha}(r) = \bar{M}_{eff}^{-1} \left[ \frac{d}{dr}
   + \bar{W}(r)\right] \bar{G}_{\alpha}(r).
 \end{eqnarray}
The Schr\"{o}dinger-equivalent effective mass and potentials are
defined as follows: for the nucleon
 \begin{eqnarray}
&&  M_{eff}= E_{\alpha}+ M_{N} - {\rm g}_{\sigma}\sigma (r)
 - {\rm g}_{\omega}\omega_{0}(r) -\frac{1}{2}{\rm
 g}_{\rho}\tau_{0\alpha}R_{0,0}(r) -\frac{1}{2}e
 \left(1+\tau_{0\alpha}\right) A_{0}(r), \label{effmn} \\
&& U_{eff}= M_{N} - {\rm g}_{\sigma}\sigma(r)
 + {\rm g}_{\omega}\omega_{0}(r)+\frac{1}{2}{\rm
 g}_{\rho}\tau_{0\alpha}R_{0,0}(r) + \frac{1}{2}e\left(
 1+\tau_{0\alpha}\right) A_{0}(r), \\
 && W(r)=\frac{\kappa_{\alpha}}{r}-\frac{f_{\omega}}{2M_{N}}\left(
    \partial_{r}\omega_{0}(r)\right)
    -\frac{f_{\rho}}{4M_{N}}\tau_{0\alpha}\left(\partial_{r}R_{0,0}(r)\right),
 \end{eqnarray}
for the anti-nucleon
 \begin{eqnarray}
&&  \bar{M}_{eff}= \bar{E}_{\alpha}+
    M_{N}-{\rm g}_{\sigma}\sigma(r)
 + {\rm g}_{\omega}\omega_{0}(r) -\frac{1}{2}{\rm
 g}_{\rho}\tau_{0\alpha}R_{0,0}(r) +\frac{1}{2}e
 \left(1+\tau_{0\alpha}\right) A_{0}(r), \\
&& \bar{U}_{eff}= M_{N} - {\rm g}_{\sigma}\sigma(r)
 - {\rm g}_{\omega}\omega_{0}(r)+\frac{1}{2}{\rm
 g}_{\rho}\tau_{0\alpha}R_{0,0}(r) - \frac{1}{2}e\left(
 1+\tau_{0\alpha}\right) A_{0}(r), \\
 && \bar{W}(r)=\frac{\kappa_{\alpha}}{r}+\frac{f_{\omega}}{2M_{N}}\left(
    \partial_{r}\omega_{0}(r)\right)
    -\frac{f_{\rho}}{4M_{N}}\tau_{0\alpha}\left(\partial_{r}R_{0,0}(r)\right).
  \label{wan}
 \end{eqnarray}
One can see that the difference between the equations of nucleons
and anti-nucleons  relies only on the definition of the effective
masses and potentials, that is, the vector fields change their
signs. The G-parity comes out automatically. The main ingredients
of the equations are meson fields which can be obtained through
solving the Laplace equations of mesons in spherical nuclei. The
source terms of the meson-field equations are  various densities
containing the contributions both from the valence nucleons and
the Dirac sea. They are evaluated by means of the derivative
expansion technique \cite{Ait84}. In numerical calculations the
Laplace equations and the equations of nucleons and anti-nucleons
are solved in an iterative procedure self-consistently to
determine the potentials and wave functions. The energy spectra of
the nucleon and the anti-nucleon are computed by means of the
following equations
\begin{eqnarray}
 E_{\alpha} &=& \int^{\infty}_{0} dr \lbrace G_{\alpha}(r) \left[
 -\frac{d}{dr} + W(r) \right] F_{\alpha}(r) + F_{\alpha}(r)
 \left[ \frac{d}{dr} + W(r) \right] G_{\alpha}(r)
 \nonumber \\
&&  + G_{\alpha}(r)U_{eff}G_{\alpha}(r) - F_{\alpha}(r) \left[
 M_{eff}- E_{\alpha} \right] F_{\alpha}(r) \rbrace, \\
 \vspace{0.4cm}
 \bar{E}_{\alpha} &=& \int^{\infty}_{0} dr \lbrace
\bar{G}_{\alpha}(r)\left[ -\frac{d}{dr} + \bar{W}(r) \right]
\bar{F}_{\alpha}(r)
 + \bar{F}_{\alpha}(r)
 \left[\frac{d}{dr} + \bar{W}(r) \right]\bar{G}_{\alpha}(r)
  \nonumber \\
&&  + \bar{G}_{\alpha}(r)\bar{U}_{eff}\bar{G}_{\alpha}(r)
 - \bar{F}_{\alpha}(r) \left[ \bar{M}_{eff}
 - \bar{E}_{\alpha} \right] \bar{F}_{\alpha}(r) \rbrace.
 \end{eqnarray}

\end{sloppypar}

\begin{center}
{\bf III. NUMERICAL RESULTS AND DISCUSSIONS}
\end{center}
\begin{sloppypar}
The parameters of the model are fixed in a least-square fit to the
properties of eight spherical nuclei. They are presented in Table
I and denoted as the RHAT set and the RHA1 set for the two cases
of with and without the tensor-coupling terms. The major result is
that a large effective nucleon mass $m^{*}/M_{N} \approx 0.78$ is
obtained. The contributions of the vacuum to the scalar density
and baryon density are depicted in Fig.~2. The computations are
performed with the RHAT set of parameters for $^{40}{\rm Ca}$.
Noticeable influence from the Dirac sea can be found for the
scalar density while the effect on the baryon density is
relatively small.

\begin{center}
\fbox{Table I}   \hspace{6cm}   \fbox{Fig.~2}
\end{center}
\begin{center}
\fbox{Table II}   \hspace{6cm}   \fbox{Table III}
\end{center}

In Table II and III we present the single-particle energies of
protons (neutrons) and anti-protons (anti-neutrons) in three
spherical nuclei of $^{16}{\rm O}$, $^{40}{\rm Ca}$ and
$^{208}{\rm Pb}$. The binding energies per nucleon and the {\em
rms} charge radii are given too. The experimental data are taken
from Ref. \cite{Mat65}. It can be found that  the relativistic
Hartree approach taking into account the vacuum effects can
reproduce the observed binding energies , {\em rms} charge radii
and particle spectra quite well. Because of the large effective
nucleon mass, the spin-orbit splitting on the $1p$ levels is
rather small in the RHA1 model. The situation has been ameliorated
conspicuously in the RHAT model incorporating the tensor couplings
for the $\omega$- and $\rho$-meson, while a large $m^{*}$ stays
unchanged. On the other hand, the anti-particle energies computed
with the RHAT set of parameters are 20 -- 30 MeV larger than that
reckoned with the RHA1 set.

\end{sloppypar}

\vspace{1cm}
 \noindent This work was supported by the National
Natural Science Foundation of China under the grant 10275072 and
the Research Fund for Returned Overseas Chinese Scholars.

\newpage
\def\baselinestretch{1.0}

\begin{table}
\caption{Parameters of the RHA models as well as the corresponding
saturation properties. $M_{N}$ and $m_{\rho}$ are fixed during the
fit.} \vspace{0.5cm}
\begin{center}
\begin{tabular}{lcc}   \\
 \hline \hline
 & RHA1 & RHAT \\
 \hline
$M_{N}$ (MeV)        &    938.000  & 938.000 \\
$m_{\sigma}$ (MeV)   &    458.000  & 450.000 \\
$m_{\omega}$ (MeV)   &    816.508  & 814.592 \\
$m_{\rho}$ (MeV)     &    763.000  & 763.000 \\
${\rm g}_{\sigma}$   &    7.1031   & 7.0899 \\
${\rm g}_{\omega}$   &   8.8496   & 9.2215 \\
${\rm g}_{\rho}$     &    10.2070  & 11.0023 \\
$b$ (fm$^{-1}$)      &    24.0870  & 18.9782  \\
$c$                  & $-$15.9936 & $-$ 27.6894 \\
$f_{\omega}/M_{N}$ (fm) &     0.0      & 2.0618   \\
$f_{\rho}/M_{N}$ (fm)   &     0.0      & 45.3318  \\
\\
$\rho_{0}$ (fm$^{-3}$)  &     0.1524  &  0.1493  \\
$E/A$ (MeV)             &    $-$16.98 & $-$ 16.76  \\
$m^{*}/M_{N}$           &    0.788   & 0.779  \\
$K$ (MeV)               &     294     & 311  \\
$a_{4}$ (MeV)           &     40.4    & 44.0  \\
\hline \hline
\end{tabular}
\end{center}
\end{table}

\begin{table}
\caption{The single-particle energies of both protons and
anti-protons as well as the binding energies per nucleon and the
{\em rms} charge radii in $^{16}{\rm O}$, $^{40}{\rm Ca}$ and
 $^{208}{\rm Pb}$. } \vspace{0.5cm}
 \begin{center}
{\small
\begin{tabular}{cccc}
\hline \hline
 & RHA1 & RHAT & Expt. \\
 \hline
$\;\;\;\;\;\;\;\;^{16}{\rm O}$  &        &      &       \\
$E/A$ (MeV)      &   8.00  & 7.94 &  7.98 \\
$r_{ch}$ (fm)    &   2.66  & 2.64 &  2.74 \\
 PROTONS         &         &      &       \\
$1s_{1/2}$ (MeV) &   30.68 & 31.63&  40$\pm$8 \\
$1p_{3/2}$ (MeV) &   15.23 & 16.18&  18.4   \\
$1p_{1/2}$ (MeV) &   13.24 & 12.22&  12.1   \\
 ANTI-PRO.       &         &      &         \\
$1\bar{s}_{1/2}$ (MeV) &   299.42 & 328.55 &     \\
$1\bar{p}_{3/2}$ (MeV) &   258.40 & 283.44 &     \\
$1\bar{p}_{1/2}$ (MeV) &   258.93 & 285.87 &     \\
\hline
$\;\;\;\;\;\;\;^{40}{\rm Ca}$ &       &      &       \\
$E/A$ (MeV)      &   8.73   &  8.62   &  8.55   \\
$r_{ch}$ (fm)    &   3.42   &  3.41   &  3.45   \\
PROTONS          &          &         &         \\
$1s_{1/2}$ (MeV) &   36.58  &  37.01  &50$\pm$11\\
$1p_{3/2}$ (MeV) &   25.32  &  25.95  &         \\
$1p_{1/2}$ (MeV) &   24.03  &  23.63  &34$\pm$6 \\
ANTI-PRO.        &             &         &         \\
$1\bar{s}_{1/2}$ (MeV)& 339.83 &  367.90 &         \\
$1\bar{p}_{3/2}$ (MeV)& 309.24 &  332.10 &         \\
$1\bar{p}_{1/2}$ (MeV)& 309.52 &  333.37 &         \\
\hline
$\;\;\;\;\;\;\;^{208}{\rm Pb}$ &        &      &       \\
$E/A$ (MeV)      &   7.93   &  7.88   &  7.87   \\
$r_{ch}$ (fm)    &   5.49   &  5.46   &  5.50   \\
 PROTONS         &         &         &         \\
$1s_{1/2}$ (MeV) &   40.80  &  41.74  &         \\
$1p_{3/2}$ (MeV) &    36.45  &  37.38  &         \\
$1p_{1/2}$ (MeV) &    36.21  &  37.18  &         \\
 ANTI-PRO.       &           &         &         \\
$1\bar{s}_{1/2}$ (MeV)&354.18  &  377.37 &         \\
$1\bar{p}_{3/2}$ (MeV)&344.48  &  366.95 &         \\
$1\bar{p}_{1/2}$ (MeV)&344.52  &  367.24 &         \\
\hline \hline
\end{tabular}
 } \end{center}
\end{table}

\begin{table}
\caption{The single-particle energies of both neutrons and
anti-neutrons.} \vspace{0.5cm}
\begin{center}
{\small
\begin{tabular}{cccc}
\hline \hline
 &  RHA1 & RHAT & Expt. \\
 \hline
$\;\;\;\;\;\;\;\;^{16}{\rm O}$  &       &      &       \\
NEUTRONS          &          &       &            \\
$1s_{1/2}$ (MeV)  &  34.71   & 35.78 &  45.7   \\
$1p_{3/2}$ (MeV)  &  19.04   & 20.18 &  21.8   \\
$1p_{1/2}$ (MeV)  &  17.05   & 15.75 &  15.7   \\
ANTI-NEU.         &         &       &         \\
$1\bar{s}_{1/2}$ (MeV)& 293.23  & 322.47&         \\
$1\bar{p}_{3/2}$ (MeV)& 252.48  & 277.94&         \\
$1\bar{p}_{1/2}$ (MeV)& 252.97  & 279.22&         \\
\hline
$\;\;\;\;\;\;\;^{40}{\rm Ca}$   &       &      &       \\
NEUTRONS          &          &       &              \\
$1s_{1/2}$ (MeV)  &  44.48   & 44.98 &         \\
$1p_{3/2}$ (MeV)  &  32.98   & 33.83 &         \\
$1p_{1/2}$ (MeV)  &  31.71   & 30.99 &         \\
ANTI-NEU.         &          &       &             \\
$1\bar{s}_{1/2}$ (MeV)& 327.96  & 355.70&         \\
$1\bar{p}_{3/2}$ (MeV)& 298.04  & 321.07&         \\
$1\bar{p}_{1/2}$ (MeV)& 298.26  & 322.15&         \\
\hline
$\;\;\;\;\;\;\;^{208}{\rm Pb}$  &       &      &       \\
NEUTRONS        &           &       &         \\
$1s_{1/2}$ (MeV)  &   47.40   & 46.70 &         \\
$1p_{3/2}$ (MeV)  &   42.66   & 42.31 &         \\
$1p_{1/2}$ (MeV)  &   42.45   & 41.64 &         \\
ANTI-NEU.         &           &       &         \\
$1\bar{s}_{1/2}$ (MeV)& 313.18  & 334.39&         \\
$1\bar{p}_{3/2}$ (MeV)& 304.61  & 325.41&         \\
$1\bar{p}_{1/2}$ (MeV)& 304.61  & 325.28&         \\
\hline \hline
\end{tabular}
} \end{center}
\end{table}

\newpage

 \begin{figure}[htbp]
 \vspace{3cm}
 \mbox{\hskip 1.0cm \psfig{file=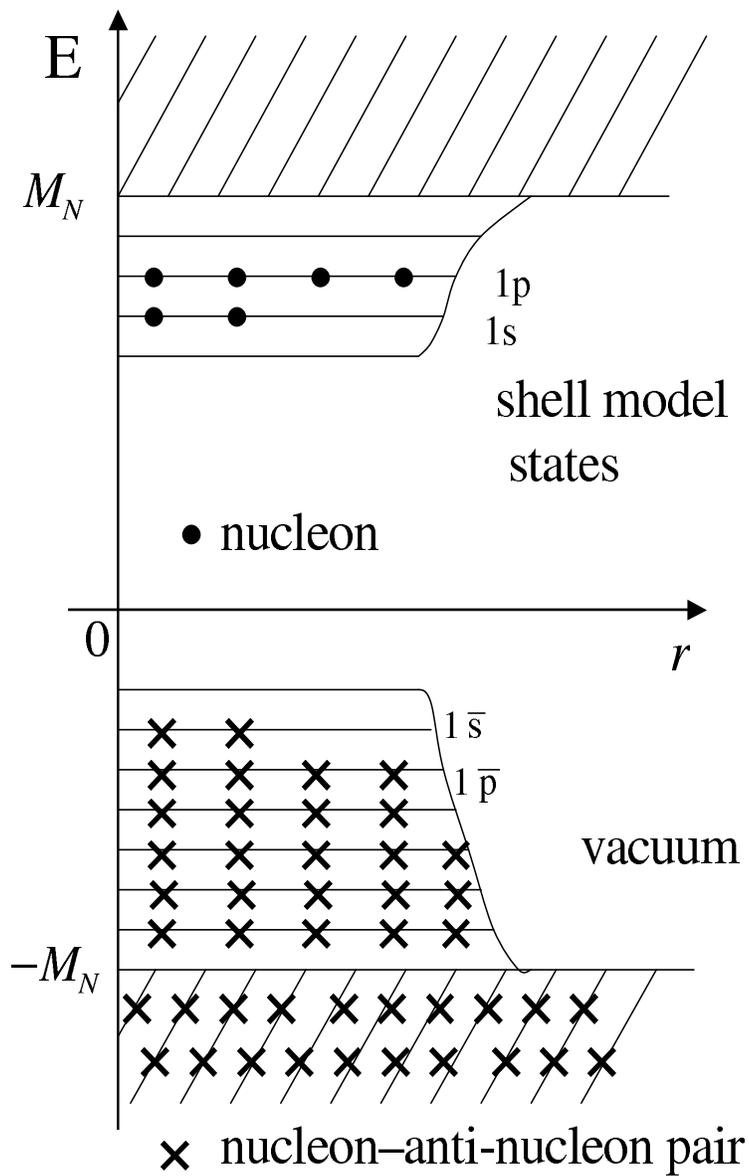,width=20cm,height=16cm,angle=-90}}
 \vspace{-0.5cm}
 \caption{A schematic picture taken from Refs. \protect{\cite{Aue86,Rei86}}
  for the energy spectra in a finite nucleus. The shell-model states are filled
  with the valence nucleons while the vacuum is full of virtual
  nucleon--anti-nucleon pairs.}
 \end{figure}

\newpage
 \begin{figure}[htbp]
 \vspace{-3.0cm}
 \mbox{\hskip -1.5cm \psfig{file=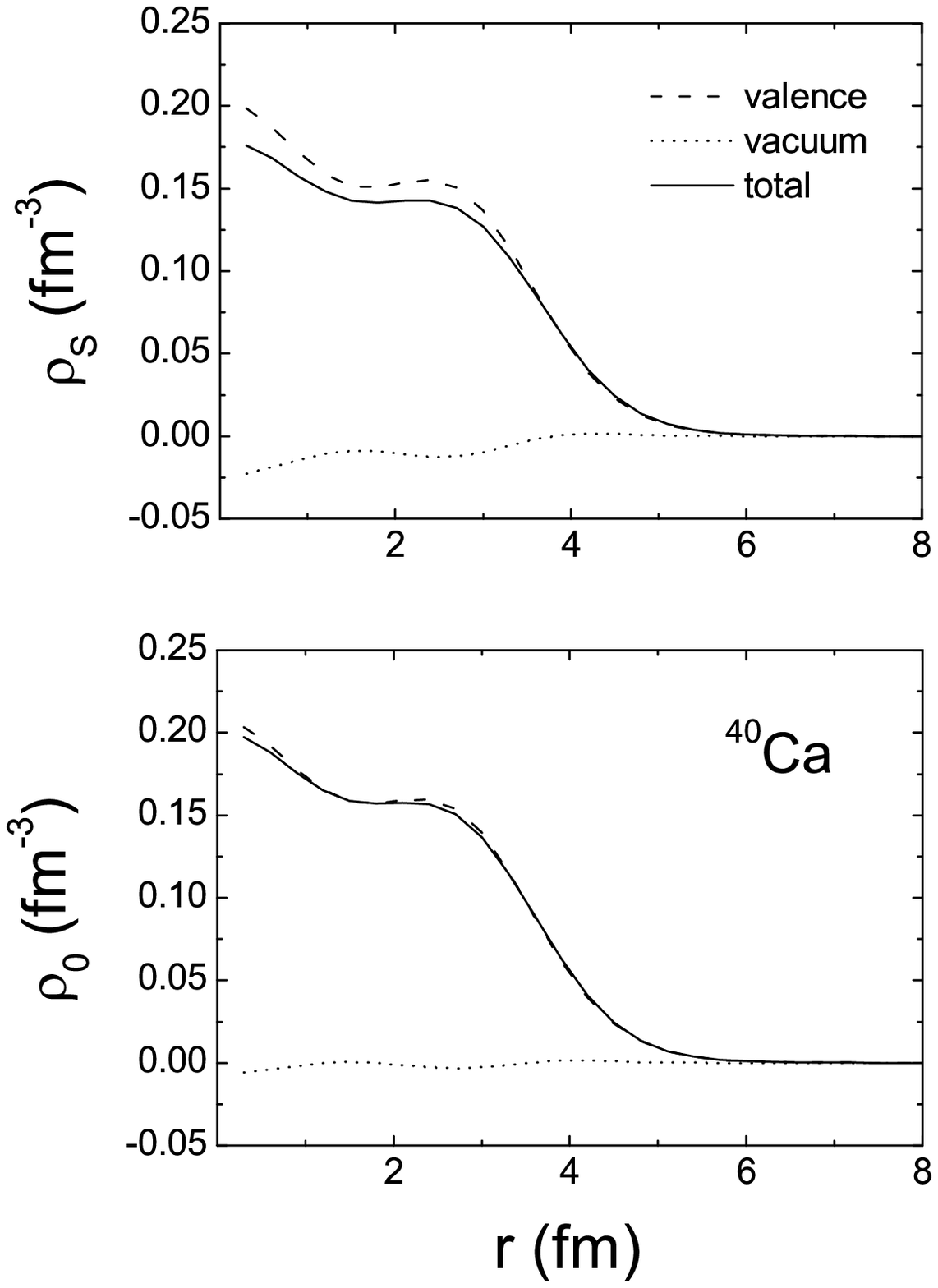,width=18cm,height=20cm,angle=0}}
 \vspace{-3.5cm}
 \caption{The scalar density and baryon density in $^{40}{\rm Ca}$. Dashed
 lines denote the contributions of valence nucleons, dotted lines represent
 the Dirac-sea effects and solid lines give the total results.}
 \end{figure}


\begin{thebibliography}{250}
\bibitem{Aue86}
   N.~Auerbach, A.S.~Goldhaber, M.B.~Johnson, L.D.~Miller, and
   A.~Picklesimer,
   Phys. Lett. {\bf B182}, 221 (1986).
\bibitem{Rei86}
   P.-G.~Reinhard, M.~Rufa, J.~Maruhn, W.~Greiner, J.~Friedrich,
   Z. Phys. {\bf A323}, 13 (1986).
\bibitem{Boh69}
   A.~Bohr and B.R.~Mottelson,
   {\it Nuclear Structure} (W.A.~Benjamin, New York, 1969).
\bibitem{Ser86}
   B.~D.~Serot and J.~D.~Walecka,
   Adv. Nucl. Phys. {\bf 16}, 1 (1986).
\bibitem{Gam90}
   Y.K.~Gambhir, P.~Ring, and A.~Thimet,
   Ann. Phys. {\bf 198}, 132 (1990).
\bibitem{Ren96}
   Zhongzhou~Ren, Z.Y.~Zhu, Y.H.~Cai, and Gongou~Xu,
   Phys. Lett. {\bf B380}, 241 (1996).
\bibitem{Men98}
   J. Meng, K. Sugawara-Tanabe, S. Yamaji, P. Ring, and A. Arima,
   Phys. Rev. {\bf C58}, R628 (1998).
\bibitem{Sug94}
   Y.~Sugahara and H.~Toki,
   Nucl. Phys. {\bf A579}, 557 (1994).
\bibitem{Hor84}
   C.J.~Horowitz and B.D.~Serot,
   Phys. Lett. {\bf B140}, 181 (1984).
\bibitem{Per86}
   R.J.~Perry,
   Phys. Lett. {\bf B182}, 269 (1986);
   Nucl. Phys. {\bf A467}, 717 (1987).
\bibitem{Was88}
   D.A.~Wasson,
   Phys. Lett. {\bf B210}, 41 (1988).
\bibitem{Fox89}
   W.R.~Fox,
   Nucl. Phys. {\bf A495}, 463 (1989).
\bibitem{Fur89}
   R.J.~Furnstahl and C.E.~Price,
   Phys. Rev. {\bf C40}, 1398 (1989);
   Phys. Rev. {\bf C41}, 1792 (1990).
\bibitem{Bea00}
    I.G.~Bearden, H.~Boggild, J.~Boissevain et al.,
    Phys. Rev. Lett. {\bf 85}, 2681 (2000).
\bibitem{Arm00}
    T.A.~Armstrong and the E864 Collaboration,
    Phys. Rev. Lett. {\bf 85}, 2685 (2000).
\bibitem{Mao99}
   G.~Mao, H.~St\"{o}cker, and W.~Greiner,
   Int. J. Mod. Phys. {\bf E8}, 389 (1999);
   AIP Conf. Proc. {\bf 597}, 112 (2001).
\bibitem{Mao03}
   G.~Mao,
   Phys. Rev. {\bf C67}, 044318 (2003);
   High Ene. Phys. Nucl. Phys. {\bf 27}, 692 (2003) (in chinese).
\bibitem{Bog77}
   J.~Boguta and A.R.~Bodmer,
   Nucl. Phys. {\bf A292}, 413 (1977).
\bibitem{Bjo64}
   J.D.~Bjorken and S.D.~Drell,
   {\em Relativistic Quantum Mechanics} (McGraw-Hill, New York, 1964).
\bibitem{Ait84}
   I.J.R.~Aitchison and C.M.~Fraser,
   Phys. Lett. {\bf B146}, 63 (1984);
   O.~Cheyette,
   Phys. Rev. Lett. {\bf 55}, 2394 (1985);
   C.M.~Fraser,
   Z. Phys. {\bf C28}, 101 (1985);
   L.H.~Chan,
   Phys. Rev. Lett. {\bf 54}, 1222 (1985).
\bibitem{Mat65}
   J.H.E.~Mattauch, W.~Thiele, and A.H.~Wapstra,
   Nucl. Phys. {\bf 67}, 1 (1965);
   D.~Vautherin and D.M.~Brink,
   Phys. Rev. {\bf C5}, 626 (1972);
   H.~de Vries, C.W.~de Jager, and C.~de Vries.
   At. Data Nucl. Data Tables {\bf 36}, 495 (1987).
\end{thebibliography}
\end{document}